\documentclass[amsmath,amssymb,twocolumn,nofootinbib]{revtex4-1}
\usepackage[dvipsnames]{xcolor}
\usepackage{graphicx,epsfig,hyperref}
\hypersetup{unicode=true,colorlinks=true,linkcolor=NavyBlue,citecolor=NavyBlue,urlcolor=black}
\usepackage[normalem]{ulem}
\begin{document}
\title{Regular rotating black hole: to Kerr or not to Kerr?}
\author{Alexander Kamenshchik}
\email{kamenshchik@bo.infn.it}
\affiliation{Dipartimento di Fisica e Astronomia, Universit\`{a} di Bologna,
via Irnerio 46, 40126 Bologna, Italy \\ I.N.F.N., Sezione di Bologna, I.S. FLAG, viale B. Pichat 6/2, 40127 Bologna, Italy}
\author{Polina Petriakova}
\email{polina.petriakova@bo.infn.it}
\affiliation{Dipartimento di Fisica e Astronomia, Universit\`{a} di Bologna,
via Irnerio 46, 40126 Bologna, Italy \\ I.N.F.N., Sezione di Bologna, I.S. FLAG, viale B. Pichat 6/2, 40127 Bologna, Italy}
\begin{abstract}
We examine the Newman{--}Janis algorithm's application to an exact regular static solution sustained by a minimally coupled scalar field with a non-standard kinetic term. Although coordinate complexification leads to a regular Kerr-like black hole, we are facing discrepancies in Einstein's equations in a fairly small domain, for which the regularizing parameter is responsible. Outside this most intriguing region, the geometry is nothing but the standard Kerr spacetime.
%\pacs{}
\end{abstract}
\maketitle
\section{Introduction}
Despite the long history of efforts at spacelike singularity resolution in the GR classical solutions, the investigation of the non-singular (or so-called regular) black holes and their regular rotating counterparts is extremely popular nowadays \cite{Lobo:2020ffi,Bonanno:2020fgp,Brahma:2020eos,Franzin:2021vnj,Maeda:2021jdc,Xu:2021lff,Mazza:2021rgq,Simpson:2021dyo,Simpson:2021zfl,Simpson:2018tsi,Bronnikov:2022bud,Zhou:2022yio,Bonanno:2022jjp,Mazza:2023iwv,Casadio:2023iqt,Zhou:2023lwc}; see also recent reviews \cite{Sebastiani:2022wbz,Torres:2022twv,Lan:2023cvz} and references therein. To construct a %spherically symmetric 
static regular solution, one relies on one of the following approaches: i) to solve Einstein’s field equations associated with a special kind of spacetime symmetry and matter sources \cite{Dymnikova:1992ux,Ayon-Beato:1998hmi,Ayon-Beato:1999kuh,Bronnikov:2000vy,Dymnikova:2001fb,Dymnikova:2003vt,Ayon-Beato:2004ywd,Bronnikov:2005gm,Bronnikov:2006fu,Smailagic:2010nv}; ii) to derive a solution as quantum corrections to the classical one \cite{Kazakov:1993ha,Bonanno:2000ep,Koch:2014cqa,Nicolini:2019irw,Ishibashi:2021kmf,Konoplya:2023aph}; or iii) to write the metric ad hoc, motivating it by phenomenological "tractability" \cite{Bardeen,Hayward:2005gi,Neves:2014aba,Simpson:2018tsi,Lobo:2020ffi,Franzin:2021vnj,Mazza:2021rgq,Simpson:2021dyo,Simpson:2021zfl}, and try to analyze the effective matter content. Nevertheless, figuring out the physical source sustaining the latter's appealing spacetimes is not trivial. In particular, one can regularize the Schwarzschild metric by some new length scale parameter $b$, %i.e., replacing radial coordinate with $r \rightarrow \sqrt{r^2+b^2}$.
resulting in a richer causal structure that seamlessly interpolates between a traversable wormhole and the so-called black-bounce geometry \cite{Simpson:2018tsi}. This, even one-parameter, extension is sustained by a combination of a minimally coupled phantom scalar field and a magnetic field within nonlinear electrodynamics \cite{Bronnikov:2021uta}. %as well as the so-called charged black-bounce spacetime \cite{Franzin:2021vnj}. %, that violate the null energy conditions. 
The physical sources for other black-bounce spacetimes, not to mention numerous regular black hole models, are unknown. The only known thing about the sources in GR is the necessity to violate the energy dominance conditions to avoid singularities \cite{Hawking:1970zqf}. Generalizing to realistic cases, rotating spacetimes, by imposing axial symmetry is more challenging. %much worse.

Apart from Kerr's unique exact solution \cite{Kerr:1963ud}, there are still only a few ways to introduce rotation into spacetime. The first in origin is the Newman{--}Janis~algorithm~(NJA). This algorithm provides a set of steps to derive the axially symmetric %(rotating) %vacuum or electrovacuum 
solutions from the spherically symmetric ones: %(static) ones:
the Kerr metric from the Schwarzschild one \cite{Newman:1965tw} and the Kerr{--}Newman from the Reissner{--}Nordstr\"{o}m~one~\cite{Newman:1965my}. This approach is based on introducing a vierbein, or tetrad, of null vectors and a series of complex coordinate conjugation transformations. %see details and generalizations of complex coordinate transformations in \cite{Erbin:2016lzq}.  
We do not dwell on the steps' details here, since we follow the NJA below. The rigorous validity of a complex coordinate transform in the Schwarzschild{--}to{--}Kerr case rests on two assumptions: the spacetime is an empty solution of Einstein's equations and belongs to the Kerr{--}Schild algebraic class~\cite{Schiffer}. The existence of the Kerr{--}Newman solution indicates that the first one is not necessary for the NJA to be successful in general. As for the second assumption, the representation of the metric in the Kerr{--}Schild form provides validity for the complex coordinate transformation in GR~\cite{Gurses:1975vu}. However, if the metric is not reducible to this form, this does not imply that the complex coordinate transformation is unfair; no one has any conception of whether it will yield reasonable results. This is established by the fulfillment of Einstein's equations\footnote{Here, we would like to note that the application of the NJA to an arbitrary non-GR spherically symmetric solution leads to pathologies possessing naked singularities \cite{Hansen:2013owa}; application to the case of GR with a minimally coupled massless/self-interacting scalar field miscarries \cite{Pirogov:2013wia,Bogush:2020lkp,Franzin:2021yvf}.}. %, which may not hold if one tries to obtain a regular solution. 
Notwithstanding, the NJA is widely used to generate regular rotating solutions \cite{Brahma:2020eos,Xu:2021lff,Mazza:2021rgq,Johannsen:2011dh,Modesto:2010rv,Caravelli:2010ff,Ghosh:2012ji,Bambi:2013ufa}, %totally ignoring 
disregarding either the source of a seed metric and/or its rotating counterpart. The uncertainty of a coordinate transform originating from different ways of getting real quantities from complex ones \cite{Drake:1998gf,Caravelli:2010ff,Erbin:2016lzq} is also worth noting. To avoid the ambiguous %Newman{--}Janis 
coordinate complexification, one can generate a stationary solution by implication of a new unknown function arising as a conformal metric multiplier \cite{Azreg-Ainou:2014aqa,Azreg-Ainou:2014pra}. The explicit form of this function is governed by the reducibility of a rotated metric to the Boyer{--}Lindquist form \cite{Boyer:1966qh} 
and the Einstein equations. Consequently, this technique assumes that the physical source is known, though one uses this approach to gain “exact" solutions, overlooking the seed metric's sources~\cite{Xu:2021lff}. A vacuum or electrovacuum axisymmetric solution can be obtained via the Ernst equation~\cite{Ernst:1967wx,Ernst:1967by}; its generalization to other non-vacuum cases is unknown. Since the regularized GR solution requires a violation of energy dominance conditions, this may lead to interpolation between a regular black hole and a wormhole. As to the wormhole case, attempts to gain an analytical stationary generalization, even for the simplest known Bronnikov{--}Ellis wormhole \cite{Bronnikov:1973fh,Ellis:1973yv}, fail \cite{Volkov:2021blw}; whereas the remaining solutions to the non-vacuum Einstein equations are known either perturbatively in the slow rotation approximation \cite{Kashargin:2007mm,Kashargin:2008pk} or numerically \cite{Kleihaus:2014dla,Chew:2016epf}.

%\newpage 
In the present paper, we examine the Newman{--}Janis algorithm's application to the regular phantom black hole \cite{Bronnikov:2005gm}. This regular static geometry either provides a wormhole or a black hole with a Schwarzschild-like causal structure but with an asymptotically de Sitter expansion instead of a singularity. The source of the seed spacetime is a minimally coupled scalar field with a non-standard kinetic term. This completely solvable example permits the matter content’s expression via geometry. %and, moreover, a quasiglobal coordinate. 
Although naive Newman{--}Janis coordinate complexification in the scalar background leads to a regular Kerr-like black hole, we are facing discrepancies in Einstein's equations in a fairly small domain, for which the regularizing parameter is responsible. Outside this most intriguing region, the energy dominance conditions are slightly violated, and the geometry is nothing but standard Kerr's spacetime.

This paper is organized as follows: Section~\ref{seed} contains a brief review of the globally regular exact solution %, the so-called regular phantom black hole 
\cite{Bronnikov:2005gm}. The regular Kerr-like spacetime is constructed from the seed static metric via the Newman{--}Janis procedure in Section~\ref{Kerr}. Section~\ref{matter} is devoted to the matter content of the obtained static and spinning solutions, and Section~\ref{discussion} to discussion.

\section{Seed regular geometry}\label{seed}
Consider the following spherically symmetric metric:
\begin{equation}\label{spherically_metric}
ds^2=A(u)\, \bigl(d x^0\bigr)^2 - \frac{d u^2}{A(u)} - r^2\bigl(u\bigr)d\Omega_2^2 ,
\end{equation}
where $d\Omega_2^2 = \bigl(d x^2\bigr)^2+\sin^2{x^2} \bigl(d x^3\bigr)^2$ is the line element on a unit sphere; the area function $r(u)$ is  regular and positive everywhere and has at least one minimum at some $u = u_\text{min}$, at which $r(u_\text{min})>0$, $r'(u_\text{min})=0$, and $r''(u_\text{min})>0$, providing the existence of two asymptotic regions with $r(u) \sim |u|$ at $u \rightarrow \pm \infty$.  
%The corresponding components of the Ricci tensor for this metric are \begin{align} \label{Ricci_00} R^0_0 & = \frac12 A'' + A' \frac{r'}{r}\, , \\  \label{Ricci_uu} R^u_u & = \frac12 A'' + A' \frac{r'}{r} + 2 A \frac{r''}{r}\, , \\  \label{Ricci_22} R^2_2 = R^3_3 & = A' \frac{r'}{r} +  A \frac{r''}{r} + A \frac{r'^2}{r^2} - \frac{1}{r^2}, \\ \label{Ricci_scalar}  R \equiv R^\mu_\mu & = A'' + 4 A' \frac{r'}{r} + 2A\left(\frac{2r''}{r} +  \frac{r'^2}{r^2} \right) - \frac{2}{r^2}, \end{align} and nontrivial components of the Einstein tensor read as follows: 
The corresponding nontrivial Einstein tensor components read as follows:
\begin{align}\label{G00}
G^0_0 &= - A' \frac{r'}{r} - 2 A \frac{r''}{r} - A \frac{r'^2}{r^2} + \frac{1}{r^2} %= T^0_0\bigl[\phi\bigr]  
\, ,
\\  \label{Guu} G^u_u &= - A' \frac{r'}{r} - A \frac{r'^2}{r^2} +  \frac{1}{r^2} %=  T^u_u\bigl[\phi\bigr] 
\, , \\
\label{G22} G^2_2 = G^3_3 & = -\frac12 A'' -  A' \frac{r'}{r} -  A \frac{r''}{r}, \end{align}
where the prime notation refers to the derivative with respect to~$u$, $\prime \equiv d / du$ and $\prime\prime \equiv d^2 / du^2$.

The exact solution of interest to us can be derived for a minimally coupled scalar field with a wide set of possible forms of the Lagrangian $L\bigl(\phi, (\phi_{,\mu})^2\bigr)$, %$L\bigl(\phi, (\partial_\mu \phi)^2\bigr)$, 
which is able to violate the null energy condition: $T^\mu_\nu k_\mu k^\nu \geq 0$ for any null vector $k^\mu$, $k_\mu k^\mu=0$.  As some examples, one can mention scalar fields with a non-standard kinetic term, such as a phantom one, $k${--}essence, or, in particular, a tachyonic field, etc. For $\phi=\phi(u)$, as for a usual minimally coupled scalar field, all these models hold equality for the stress-energy tensor's components\footnote{An appropriate algebraic structure of the stress-energy tensor was first discussed by Gliner: to generate the regular spherically symmetric solution, the structure has the characteristics $[(1111)]$, $[11(11)]$, $[(11)(11)]$, or $[1(111)]$, where brackets denote the components' equality \cite{Gliner}. The latter one corresponds to the case we are considering here.}, $T^0_0\bigl[\phi\bigr]=T^2_2\bigl[\phi\bigr]=T^3_3\bigl[\phi\bigr]$, as well as the corresponding components of the Einstein tensor $G^\mu_\mu$ and of the Ricci tensor $R^\mu_\mu$. Using this fact, or by noting the difference between $G^0_0$ and $G^2_2$, one~obtains 
\begin{equation}\label{G_002233_equal}
\frac12 A'' - A \frac{r''}{r}  - A \frac{r'^2}{r^2} + \frac{1}{r^2} =0.
\end{equation}

To find a globally regular geometry, we choose the simplest possible area function %$r(u)=\sqrt{u^2+b^2}$, where $b=\text{const}>0$, 
\begin{equation}
 r(u)=\sqrt{u^2+b^2}, \quad b=\text{const}>0,   
\end{equation}
which is suitable for the conditions of the regular minimum and gives us the following analytic solution to the equation above\footnote{We denote $\arctan{\frac{u}{b}}$ as $\tan^{-1}{\frac{u}{b}}$ and $\text{arccot}\frac{u}{b}$ as $\cot^{-1}{\frac{u}{b}}$}:
\begin{equation}\label{A(u)c1c2}
A(u)=1+ c_1 \bigl(u^2+b^2\bigr) +c_2 \Bigl(\bigl(u^2+b^2 \bigr)\tan^{-1}{\frac{u}{b}} + ub \Bigr).
\end{equation}
Depending on $c_1$ and $c_2$ values, the obtained solution may be asymptotically flat or anti{--}de Sitter in the static region and asymptotically de Sitter in the nonstatic region. %, where the coordinates change their temporal and spatial assignments. 
By setting $c_1=- \pi c_2/2$ and $c_2b^3 \equiv u_0$ to ensure the regularity of $A(u)$ at $b \rightarrow 0$ and the Schwarzschild form, i.e., $A(u) \simeq 1 - 2u_0/3u$ at $u \rightarrow + \infty$ correspondingly, we get 
\begin{equation}\label{A(u)}
A(u)  = 1 - \frac{u_0}{b^3} \left(\bigl(u^2+b^2 \bigr) \cot^{-1}{\frac{u}{b}} - ub \right).
\end{equation}
This redshift metric function either provides a wormhole or a regular black hole with a Schwarzschild-like causal structure but with an asymptotically de Sitter expansion instead of a singularity. %, the so-called black universe \cite{Bronnikov:2006fu}. 

We have a traversable wormhole if $2b >\pi u_0$, a regular black hole if $0 < 2b < \pi u_0$ with a single horizon
at $u_h$, which is a regular zero of $A(u_h)=0$, or a regular black hole with a single extremal horizon (black throat)
at $u = 0$ if $ 2b = \pi u_0$. Beyond the event horizon (if it exists), there is a bounce to anisotropic Kantowski{--}Sachs cosmology with two scale factors, $A(u)$ and $r(u)$. In the exceptional case, by setting the constant $u_0 = 0$, one gets the Bronnikov{--}Ellis-like wormhole \cite{Bronnikov:1973fh,Ellis:1973yv}, i.e., $A(u) = 1$. Some additional features of this globally regular solution were also revealed in \cite{Chataignier:2022yic}.

\section{From the regular Schwarzschild-like black hole to the regular Kerr-like black hole}\label{Kerr}

As was mentioned in the Introduction, the Newman{--}Janis algorithm provides the derivation of the rotating solutions from the static ones. This approach is based on introducing a %tetrad, or 
vierbein of null vectors %$\{l^\mu, n^\mu, m^\mu, \bar{m}^\mu\}$ 
$\textbf{e}_\alpha = \bigl(\textbf{l},\textbf{n}, \textbf{m}, \bar{\textbf{m}}\bigr)$ and a series of complex conjugation transformations; here and in the following, the bar marks complex conjugation. In this section, we apply it to the seed regular geometry obtained above.

As the NJA's first step, we switch to a null coordinate system, replacing in \eqref{spherically_metric} the time coordinate $x^0$ with the null time coordinate $\tau$ via $ dx^0 \rightarrow d\tau = dx^0-du\, /A(u)$. This provides the metric form in the so-called Eddington{--}Finkelstein type coordinates,
\begin{equation}\label{metricEFcoord}
ds^2 = A(u) d\tau^2 + 2 d\tau du - r^2(u)d\Omega_2^2,
\end{equation}
which can be expressed in terms of the null Newman{--}Penrose tetrad formalism %\cite{Newman:1961qr} 
as 
\begin{equation}
ds^2 = \bigl( l_\mu n_\nu - m_\mu \bar{m}_\nu  \bigr)dx^\mu dx^\nu,
\end{equation}
with 
\begin{equation}\label{old_tetrad} \begin{split}
l^\mu &= \delta^\mu_u, \quad  \,\, n^\mu =\delta^\mu_\tau - \frac{A(u)}{2} \delta^\mu_u, \\ m^\mu &= \frac{1}{\sqrt{2}r(u)} \biggl( \delta^\mu_2 + \frac{i}{\sin{x^2}}\,\delta^\mu_3 \biggr), \end{split} \end{equation}
with $l^\mu$ and $n^\mu$ being real null vectors, and $m^\mu$ and its complex conjugate $\bar{m}^\mu$ being complex null vectors. They are required to satisfy the orthogonality conditions, %$l_\mu m^\mu = l_\mu \bar{m}^\mu =  n_\mu m^\mu = n_\mu \bar{m}^\mu =0$, 
the essentials for the null vectors, %$l_\mu l^\mu= n_\mu n^\mu = m_\mu m^\mu = \bar{m}_\mu \bar{m}^\mu=0$, 
and the normalization conditions; i.e., all products of these four vectors vanish, with two exceptions: $l_\mu n^\mu= - m_\mu\bar{m}^\mu =1$. 

Next, according to the Newman{--}Janis algorithm, we need to complexify the seed metric functions, i.e., replace the redshift function $A(u)$ by a new one $A(u,\bar{u})$, as
\begin{align}
A(u,\bar{u}) & = 1  +\frac{u_0}{2b^2}\bigl(u+\bar{u}\bigr) \, - \\ \nonumber & \,\, - \frac{u_0}{2b^3}\biggl(\bigl(u^2+b^2\bigr)\cot^{-1}{\frac{u}{b}} + \bigl(\bar{u}^2+b^2\bigr)\cot^{-1}{\frac{\bar{u}}{b}}\biggr),
\end{align}
the area function $r(u)$ in the pair of complex null vectors $m^\mu$ and $\bar{m}^\mu$ as
\begin{equation}
r(u)=\sqrt{u^2+b^2\,}, \quad \bar{r}(u)= \sqrt{\bar{u}^2+b^2\,},
\end{equation}
requiring at $u=\bar{u}$ the recovery of initial vierbein~\eqref{old_tetrad}, and apply complex transformation coordinates 
\begin{equation}\label{coord_transform}
x^\mu \rightarrow x^{\prime \mu} = x^\mu - i a \cos{x^2} \bigl( \delta^\mu_{\tau} - \delta^\mu_{u} \bigr),
\end{equation}
treating the primed coordinates as real. Through the null complex tetrad transform, $\text{e}^{\mu}_\alpha \rightarrow \text{e}^{\prime \mu}_\alpha = \text{e}^\nu_\alpha \, \partial x^{\prime \mu}/ \partial x^\nu$, and the use of the ensuing new light-like vectors 
\begin{equation} \begin{split}
l^{\, \prime \mu} &%\equiv \frac{\partial x^{\prime \mu}}{\partial x^\nu} l^\nu 
= \delta^\mu_{u'}, \qquad n^{\prime \mu} = \delta^\mu_{\tau'} - \frac{A\bigl(u',{x^2}'\bigr)}{2} \delta^\mu_{u'}, \\
m^{\prime \mu} &= \frac{\,\,\,\, ia\sin{{x^2}'}\Bigl( \delta^\mu_{\tau'} - \delta^\mu_{u'} \Bigr) + \delta^\mu_{2'} + \cfrac{i}{\sin{{x^2}'}} \,\, \delta^\mu_{3'}}{\sqrt{2}\, \bar{r}\bigl(u',{x^2}'\bigr)}, \\
\bar{m}'^\mu &= \frac{-ia\sin{{x^2}'}\Bigl( \delta^\mu_{\tau'} - \delta^\mu_{u'} \Bigr) + \delta^\mu_{2'} - \cfrac{i}{\sin{{x^2}'}} \,\, \delta^\mu_{3'}}{\sqrt{2}\, r\bigl(u',{x^2}'\bigr)}, \\
&
\end{split} \end{equation}
it yields the new %${g'}^{\mu\nu} = {l'}^\mu {n'}^\nu + {l'}^\nu {n'}^\mu - {m'}^\mu \bar{m}'^\nu - {m'}^\nu \bar{m}'^\mu$ 
$g^{\, \prime \mu\nu} = 2l^{\,\prime (\mu} n^{\prime \, \nu)}-2m^{\prime (\mu}{\bar{m}}^{\prime \, \nu)}$ 
expression,
\begin{widetext} \begin{equation} \label{metric_prime}  {g^\prime}^{\mu\nu} = \small \begin{pmatrix}   -\dfrac{a^2\sin^2{{x^2}'}}{r\bar{r}(u',{x^2}'\bigr)} & 1 + \dfrac{a^2\sin^2{{x^2}'}}{r\bar{r}(u',{x^2}'\bigr)} & 0 & -\dfrac{a}{r\bar{r}(u',{x^2}'\bigr)} \\ 1 + \dfrac{a^2\sin^2{{x^2}'}}{r\bar{r}(u',{x^2}'\bigr)} & -A(u',{x^2}'\bigr) - \dfrac{a^2\sin^2{{x^2}'}}{r\bar{r}(u',{x^2}'\bigr)} & 0 & \dfrac{a}{r\bar{r}(u',{x^2}'\bigr)} \\ 0 & 0 &  -\dfrac{1}{r\bar{r}(u',{x^2}'\bigr)} & 0 \\ -\dfrac{a}{r\bar{r}(u',{x^2}'\bigr)} &  \dfrac{a}{r\bar{r}(u',{x^2}'\bigr)} & 0 &  -\dfrac{1}{r\bar{r}(u',{x^2}'\bigr)\sin^2{{x^2}'}} \\ \end{pmatrix}, \normalsize \end{equation} %\end{widetext}
and the inverse metric in the ingoing Eddington{--}Finkelstein coordinates %, as in the Kerr original solution, 
%\begin{widetext} \begin{equation} \label{metric_prime_Kerr_like_coord} g^\prime_{\mu\nu}  = \small \begin{pmatrix} A\bigl(u',{x^2}'\bigr) & 1 & 0 & a\Bigl(1-A\bigl(u',{x^2}'\bigr)\Bigr)\sin^2{{x^2}'} \\  1 & 0 & 0 & -a\sin^2{{x^2}'} \\  0 & 0 & -r\bar{r}\bigl(u',{x^2}'\bigr) & 0 \\  a\Bigl(1-A\bigl(u',{x^2}'\bigr)\Bigr)\sin^2{{x^2}'} &  -a\sin^2{{x^2}'}  & 0 & -\biggl(a^2\Bigl(2-A\bigl(u',{x^2}'\bigr)\Bigr)\sin^2{{x^2}'} + r\bar{r}\bigl(u',{x^2}'\bigr)\biggr)\sin^2{{x^2}'}\\ \end{pmatrix}, \normalsize \end{equation}
being written via a line element is
\begin{equation} \begin{gathered}\label{metric_prime_Kerr_like_coord}  %{ds'}^2= A\bigl(u',{x^2}'\bigr)\bigl(d\tau'\bigr)^2 + 2d\tau'du' + 2a\Bigl(1- A\bigl(u',{x^2}' \, \bigr) \Bigr) \sin^2{{x^2}'} d\tau'd{x^3}' - 2a\sin^2{{x^2}'}du'd{x^3}' - \\ - r\bar{r} \bigl(d{x^2}'\bigr)^2 - \Bigl(a^2\Bigl(2-A\bigl(u',{x^2}'\, \bigr)\Bigr)\sin^2{{x^2}'} + r\bar{r}\Bigr)\sin^2{{x^2}'} \bigl(d{x^3}'\bigr)^2. 
{ds'}^2= A\bigl(u',{x^2}'\bigr)\Bigl(d\tau'-a\sin^2{{x^2}'}d{x^3}'\Bigr)^2 + 2 \Bigl(d\tau' - a\sin^2{{x^2}'}d{x^3}'\Bigr)\Bigl(du' + a\sin^2{{x^2}'}d{x^3}'\Bigr) \\ - r\bar{r}\bigl(u',{x^2}'\bigr) \Bigl(\bigl(d{x^2}'\bigr)^2 + \sin^2{{x^2}'} \bigl(d{x^3}'\bigr)^2\Bigr), \end{gathered} \end{equation}
where
\begin{equation}\begin{gathered}\label{A_complex} A\bigl(u',{x^2}'\bigr) = 1+\frac{u_0u'}{b^2} +  \frac{u_0 }{2b^3}\Bigl(u'^2-a^2\cos^2{{x^2}'}+b^2\Bigr)\biggl(\tan^{-1}{\frac{u'}{b+a\cos{{x^2}'}}} + \tan^{-1}{\frac{u'}{b- a\cos{{x^2}'}}}-\pi\biggr)  \\ +  \frac{au_0u'}{2b^3} \, \cos{{x^2}'} \, \ln{\frac{{u'}^2+\bigl(b-a\cos{{x^2}'}\bigr)^2}{{u'}^2+\bigl(b+a\cos{{x^2}'}\bigr)^2}} \end{gathered} \end{equation} and 
\begin{equation}\label{r_rBar}
r\bar{r} \bigl(u',{x^2}'\bigr) 
= \sqrt{\bigl(u'^2-a^2\cos^2{{x^2}'}+b^2\bigr)^2 + 4a^2u'^2\cos^2{{x^2}'}}.
\end{equation}

The obtained geometry
\eqref{metric_prime_Kerr_like_coord} does not contain Kerr's usual ring coordinate singularity at $u'=0$ and ${x^2}'=\pi/2$, %,see $g_{\tau \tau}$ and \eqref{A_complex}. 
and it turns into the Kerr original one \cite{Kerr:1963ud} at the $b\rightarrow 0$ limit:
\begin{equation}\label{metric_prime_Kerr_like_coord_limit}
\left. g^\prime_{\mu\nu}\right|_{b = 0} = \small
\begin{pmatrix} 1-\cfrac{2u_0u'}{3\bigl({u'}^2+a^2\cos^2{{x^2}'}\bigr)} & 1 & 0 & \cfrac{2au_0 u'\sin^2{{x^2}'}}{3\bigl({u'}^2+a^2\cos^2{{x^2}'}\bigr)} \\ 
1 & 0 & 0 & -a\sin^2{{x^2}'} \\ 
0 & 0 & -{u'}^2-a^2\cos^2{{x^2}'}& 0 \\ 
\cfrac{2au_0 u'\sin^2{{x^2}'}}{3\bigl({u'}^2+a^2\cos^2{{x^2}'}\bigr)} &  -a\sin^2{{x^2}'}  & 0 & -\Biggl({u'}^2 + a^2 + \cfrac{2a^2 u_0 u' \sin^2{{x^2}'}}{3\bigl({u'}^2+a^2\cos^2{{x^2}'}\bigr)}\Biggr)\sin^2{{x^2}'} \\ \end{pmatrix}, \normalsize
\end{equation} \normalsize \end{widetext}
and, as expected, the Schwarzschild one in the Eddington{--}Finkelstein null coordinates for $a=0$. %This spacetime \eqref{metric_prime_Kerr_like_coord} is algebraically general, i.e., only the Weyl scalar$\Psi_0 = C_{\sigma \rho \gamma \delta}l^\sigma m^\rho l^\gamma m^\delta$ is equal to zero. %Note that putting $u_0/3=0$, we have 

The curvature invariants for the obtained rotated solution \eqref{metric_prime_Kerr_like_coord} are finite in the entire range of the $u'$ coordinate. The only potentially dangerous denominator of curvature invariants can arise at the value $u'=0$. Fortunately, due to the structure of $r\bar{r} \bigl(u',{x^2}'\bigr)$, see \eqref{r_rBar},
the Ricci scalar, the Ricci tensor squared, and the Kretschmann scalar, %(the Riemann tensor squared)
\begin{equation}\begin{gathered}
 R \sim \bigl(r\bar{r}\bigr)^{-3}, \quad R_{\alpha\beta}R^{\alpha\beta}  \sim \bigl(r\bar{r}\bigr)^{-6}, \\ \text{and} \quad \mathcal{K} \equiv  R_{\alpha\beta\gamma\delta}R^{\alpha\beta\gamma\delta} \sim \bigl(r\bar{r}\bigr)^{-6},    
\end{gathered}
\end{equation}
are globally regular %as is the Weyl contraction, $C_{\alpha\beta\gamma\delta}C^{\alpha\beta\gamma\delta} \equiv R_{\alpha\beta\gamma\delta}R^{\alpha\beta\gamma\delta} - 2R_{\alpha\beta}R^{\alpha\beta} + R^2/3$, (in the sense of Bardeen \cite{Bardeen}) 
if $a \neq b$, and at the $b \rightarrow 0$ limit the standard features of the Kerr spacetime are observed. %If zero of the obtained metric function \eqref{A_complex} lies before the minimum of the area function \eqref{r_rBar}, i.e., at $u'>0$, by resolving the transcendental equation, the ergoregion and the event horizon can be determined.

As the final step of the NJA, one reverts the metric to the Boyer{--}Lindquist coordinates, which furnish only a single off-diagonal component, $g_{\tau {x}^3}$. Nevertheless, it is not always possible to find such integrable coordinate transformation,
$d\tau' \rightarrow d\tau=d\tau' - \alpha(u') du'$ %, $du=du'$, $d{x}^2= d{x'}^2$, 
and $d{x'}^3 \rightarrow d{x}^3 = d{x'}^3 - \beta(u') du'$, that preserve $\alpha(u')$ and $\beta(u')$ independent of ${x^2}'$ as in our non-empty case here. This~is guaranteed only if spacetime can be thrown into the Kerr{--}Schild form that corresponds  to algebraically special classes belonging to the %empty solutions' 
Petrov classification \cite{Chandrasekhar:1985kt}. Though we note that for a small regularizing parameter~$b$, these functions 
\begin{equation}\label{alpha_beta_b_small}\begin{split}
\small \alpha(u',{x^2}') &= \frac{{g'}^{\tau u}}{{g'}^{uu}} \simeq  \frac{{u'}^2+a^2}{{u'}^2+a^2-2u_0u'/3} + O(b^2), \\ \beta(u',{x^2}') &= \frac{{g'}^{u x^3}}{{g'}^{uu}} \simeq  \frac{a}{{u'}^2+a^2-2u_0u'/3} + O(b^2),
\end{split} \end{equation} 
\normalsize provide the well-known Boyer{--}Lindquist transform \cite{Boyer:1966qh}, and spacetime \eqref{metric_prime_Kerr_like_coord}, being algebraically general, degenerates to an algebraically 
special and of Petrov type D up to $O(b^2)$; and in the slow rotation approximation,
\begin{equation} \begin{split} 
\alpha(u',{x^2}') & \simeq  \frac{1}{A(u')} - \frac{a^2\bigl(1-A(u')\bigr)}{A^2(u')r^2(u')} + O(a^4),
\\ \beta(u',{x^2}') & \simeq \cfrac{a}{A(u')r^2(u')}+ O(a^3), 
\end{split} \end{equation}
metric \eqref{metric_prime_Kerr_like_coord} can also be reduced to the Boyer{--}Lindquist representation:
\begin{widetext}
\begin{equation} \begin{gathered}
\small ds^2_{slow} \simeq \Bigl(A(u') + O\bigl(a^2\bigr)\Bigr)d\tau^2 + \Bigl(2a\sin^2{{x^2}'}\bigl(1-A(u')\bigr)+O\bigl(a^3\bigr)\Bigr)d\tau dx^3 - \Bigl(A^{-1}(u')+O\bigl(a^2\bigr)\Bigr)d{u'}^2  \\ - \Bigl(r^2(u')+ O\bigl(a^2\bigr) \Bigr)   \Bigl(\bigl(d{x^2}'\bigr)^2 + \sin^2{{x^2}'} \bigl(d{x^3}\bigr)^2\Bigr), \normalsize
\end{gathered} \end{equation} \end{widetext}
coinciding at $u' \rightarrow + \infty$ with the slow rotation limit of the standard Kerr solution %up to $O\bigl({u'}^{-3}\bigr)$. 
in these coordinates.

\section{Scalar field}\label{matter}
In this section, we consider the matter content that sustains the static seed geometry. For the simplest option mentioned in Section II, for a phantom scalar field, everything can be explicitly expressed via geometry. %and, moreover, in terms of the quasiglobal radial coordinate~$u$. 
Hereafter, we apply a series of complex conjugation transformations to the scalar field in the Newman{--}Janis spirit.

The scalar field's stress-energy tensor is 
\begin{equation}\label{SET_scalar}
%T^\phi_{\mu\nu} = \epsilon \partial_{\mu}\phi \, \partial_{\nu}\phi - \frac{\epsilon}{2} \, g_{\mu \nu}  \partial^{\alpha}\phi \, \partial_{\alpha}\phi + g_{\mu \nu} V\bigl( \phi \bigr) \, ,
%T^\mu_\nu[\phi] = \epsilon \partial^{\mu}\phi \, \partial_{\nu}\phi - \frac{\delta^\mu_\nu}{2} \, \epsilon \, \partial^{\alpha}\phi \, \partial_{\alpha}\phi + \delta^\mu_\nu V\bigl( \phi \bigr),
T^\mu_\nu[\phi] = \epsilon \phi^{,\mu} \, \phi_{,\nu} - \frac{\delta^\mu_\nu}{2} \, \epsilon \,\phi^{,\alpha} \, \phi_{,\alpha} + \delta^\mu_\nu V\bigl( \phi \bigr),
\end{equation}
where $\epsilon=+1$ corresponds to a canonical scalar field $\phi_{\text{c}}$ and $\epsilon=-1$ to a phantom one $\phi_{\text{ph}}$. %i.e., with a “wrong” sign in front of the kinetic term. 
\newpage   
As for the non-rotation case, assuming $\phi = \phi\bigl(u\bigr) $, the difference between\footnote{ $G^0_0$ and  $G^u_u$ are to be taken from \eqref{G00} and \eqref{Guu}, and $8\pi G=c=1$.} $G^\tau_\tau (=G^0_0)$ and $G^u_u$ for metric \eqref{metricEFcoord}, or $G^\tau_u$ itself, being matched with the stress-energy tensor of the scalar field \eqref{SET_scalar}, yields
\begin{equation}\label{scalar}
%G^\tau_\tau - G^u_u = G^\tau_u = 
-2\frac{r''}{r} = \epsilon {\phi'}^2 \quad \rightarrow \quad \phi_{\text{ph}}(u) = \pm \sqrt{2} \tan^{-1}{\frac{u}{b}} + \phi_0.
\end{equation}
Without loss of generality, we chose the minus sign and $\phi_0=\pi/\sqrt{2}$, resulting in $\phi_{\text{ph}}(u) = \sqrt{2} \cot^{-1}{\bigl(u / b \bigr)}$.  
The sum of $G^\tau_\tau$ and $G^u_u$ components leads to an expression for potential in terms of the quasiglobal radial coordinate~$u$:
\begin{equation}\label{V(u)}
V\bigl(u\bigr) %= -\frac{1}{r^2} \Bigl(A'r'r+A\bigl(r''r+{r'}^2\bigr)-1\Bigr) 
=\frac{u_0\Bigl(\bigl(3u^2+b^2\bigr) \cot^{-1}{\dfrac{u}{b}} -3ub\Bigr)}{b^3\bigl(u^2+b^2\bigr)}\, .
\end{equation}
Further, one can reconstruct the exact expression for the potential via the phantom scalar field:  
\begin{equation}
V\bigl(\phi_{\text{ph}}\bigr)  = \frac{u_0\phi_{\text{ph}}}{\sqrt{2}b^3} \Bigl(3- 2\sin^2{\frac{\phi_{\text{ph}}}{\sqrt{2}}}\Bigr) -  \frac{3u_0}{2b^3}\sin{\sqrt{2}\phi_{\text{ph}}}, \end{equation}
which comes from %\eqref{scalar_cot} after inverting, i.e., $u=b\cot{\cfrac{\phi_{\text{ph}}}{\sqrt{2}}}$, $\phi_{\text{ph}} \in \left(0; \pi \right)$, and \eqref{V(u)}. %$\phi_{\text{ph}} \in \left(-\pi/2; \pi/2 \right)$, and 
\eqref{V(u)} after inverting \eqref{scalar}, 
$u=b\cot{\cfrac{\phi_{\text{ph}}}{\sqrt{2}}}$. 

As for the rotating case, we intend to complexify the scalar field \eqref{scalar}, the potential \eqref{V(u)}, or the Lagrangian density itself, 
%\begin{equation}\label{L(u)}  L(u) =\frac{u_0u\bigl(3u^2+4b^2 \bigr)+b^4}{b^2\bigl(u^2+b^2\bigr)^2} -  \frac{u_0\bigl(3u^2+2b^2 \bigr)}{b^3\bigl(u^2+b^2\bigr)}\cot^{-1}{\cfrac{u}{b}}  = \frac{- u_0\bigl(3u^4+5u^2b^2+2b^4 \bigr)\cot^{-1}{\cfrac{u}{b}}+u_0ub\bigl(3u^2+4b^2 \bigr)+b^5}{b^3\bigl(u^2+b^2\bigr)^2} = \frac{b^2+u_0u}{\bigl(u^2+b^2\bigr)^2} - \cfrac{u_0\bigl(3u^2+2b^2\bigr)}{b^3\bigl(u^2+b^2\bigr)} \cot^{-1}{\cfrac{u}{b}} + \frac{3u_0u}{b^2\bigl(u^2+b^2\bigr)}, \end{equation}
i.e., replace it, as in the previous section, with new ones $\phi_{\text{ph}}(u,\bar{u})$, $V(u,\bar{u})$, and $L(u,\bar{u})$ and apply the complex transformation coordinates $x^\mu \rightarrow x^{\prime \mu}$, see~\eqref{coord_transform}. Then, for the scalar field \eqref{scalar}, dropping coordinate prime indices, we have
\begin{equation}\label{scalar_rotated} \begin{gathered} \small \phi_{\text{ph}} \bigl(u,x^2\bigr) = \frac{\pi}{\sqrt{2}} - \frac{1}{\sqrt{2}}\tan^{-1}{\dfrac{u}{b+a\cos{x^2}}} \\ \qquad \qquad  - \frac{1}{\sqrt{2}} \tan^{-1}{\dfrac{u}{b- a\cos{x^2}}}  \, , \end{gathered} \end{equation} and the stress-energy tensor \eqref{SET_scalar} appears to be
\begin{widetext} \begin{equation} \label{SET_rotation} \small T^\mu_\nu
%[\phi_{\text{ph}}]  
= \begin{pmatrix} - L\bigl(u, x^2\bigr)  & -\left( 1 + \cfrac{a^2\sin^2{x^2}}{r\bar{r}(u,x^2\bigr)} \right) {\bigl( \phi_{\text{ph}} \bigr)'_u}^2 & -\left( 1 + \cfrac{a^2\sin^2{x^2}}{r\bar{r}(u,x^2\bigr)} \right) \bigl( \phi_{\text{ph}} \bigr)'_u \bigl( \phi_{\text{ph}} \bigr)'_{x^2} & 0 \\ 0 & \left( A(u,x^2\bigr) + \cfrac{a^2\sin^2{x^2}}{r\bar{r}(u,x^2\bigr)} \right) {\bigl( \phi_{\text{ph}} \bigr)'_u}^2 - L\bigl(u, x^2\bigr) &  \left( A(u,x^2\bigr) + \cfrac{a^2\sin^2{x^2}}{r\bar{r}(u,x^2\bigr)} \right) \bigl( \phi_{\text{ph}} \bigr)'_u \bigl( \phi_{\text{ph}} \bigr)'_{x^2} & 0 \\ 0 & \cfrac{\bigl( \phi_{\text{ph}} \bigr)'_u \bigl( \phi_{\text{ph}} \bigr)'_{x^2}}{r\bar{r}(u,x^2\bigr)} & \cfrac{{\bigl( \phi_{\text{ph}} \bigr)'_{x^2}}^2}{r\bar{r}(u,x^2\bigr)} - L\bigl(u, x^2\bigr) & 0 \\  0 & - \cfrac{a{\bigl( \phi_{\text{ph}} \bigr)'_u}^2}{r\bar{r}(u,x^2\bigr)} & -\cfrac{a\bigl( \phi_{\text{ph}} \bigr)'_u \bigl( \phi_{\text{ph}} \bigr)'_{x^2}}{r\bar{r}(u,x^2\bigr)} & - L\bigl(u, x^2\bigr) \\ \end{pmatrix}, \normalsize \end{equation}
where 
\begin{align}
\small  L\bigl(u,x^2\bigr) = \frac{1}{b^2\bigl(r\bar{r}(u,x^2)\bigr)^4}\biggl( \bigl(u^2 +& b^2\bigr)^2\bigl(3u_0 u^3 + 4u_0b^2 u + b^4\bigr)+\Bigl(u_0 u\bigl(9u^4+4b^2u^2-b^4\bigr) - 2b^4\bigl(3u^2 + b^2\bigr) \Bigr)a^2\cos^2{x^2}  \nonumber \\
+\, \bigl(9u_0 u^3 - 6u_0 b^2u + b^4&\bigr)a^4\cos^4{x^2} + 3u_0 u\, a^6 \cos^6{x^2}\biggr) + \frac{u_0 u \, a \cos{x^2}}{2b\bigl(r\bar{r}(u,x^2)\bigr)^2} \, \ln{\frac{u^2+\bigl(b-a\cos{x^2}\bigr)^2}{u^2+\bigl(b+a\cos{x^2}\bigr)^2}} \, \,  \\ \nonumber + \frac{u_0}{2b^3\bigl(r\bar{r}(u,x^2)\bigr)^2} \Bigl(\bigl(3u^2 + 2b^2\bigr)\bigl(u^2 + &b^2\bigr) +  \bigl(6u^2 - 5b^2\bigr)a^2\cos^2{x^2} + 3a^4\cos^4{x^2}\Bigr)\biggl(\tan^{-1}{\frac{u}{b+a\cos{{x^2}}}} + \tan^{-1}{\frac{u}{b- a\cos{x^2}}}-\pi\biggr). \end{align}
\end{widetext} \normalsize 
The non-trivial components of the obtained stress-energy tensor \eqref{SET_rotation} are asymtotically trivial, see \eqref{scalar_rotated} at~$u \rightarrow + \infty$, behave as $T^\mu_\nu[\phi_{\text{ph}}] \sim O(b^2)$ at the $b \rightarrow 0$ limit, and turn out coinciding with the exact non-rotation ones if $a=0$. %or $x^2=\pi/2$. 

However, the mixed Einstein tensor's components $G^\mu_\nu$, which correspond to the metric \eqref{metric_prime_Kerr_like_coord}, %{--}\eqref{r_rBar}  
are all non-trivial\footnote{Howbeit, in the slow rotation limit $G^\tau_{3} \sim G^u_{\tau} \sim G^u_{3} \sim G^2_{\tau} \sim G^{2}_{3} \sim G^3_{\tau} \sim O(a^3)$, while other non-trivial components $\sim O(a^2)$.}. The Einstein equations, $G^\mu_\nu= T^\mu_\nu[\phi_{\text{ph}}]$, ensuing from the made coordinate transformations are satisfied asymptotically, being noticeably violated only at distances on the order of the regularization parameter $b$. Regarding the Newman{--}Janis algorithm in the scalar background, a similar observation was recently made in \cite{Makukov:2023den}. We do not present here quite cumbersome expressions for the Einstein tensor's components, but one can verify that $G^\mu_\nu \sim O(b^2)$ at the $b \rightarrow 0$ limit via Maple, etc. Since our constructed geometry \eqref{metric_prime_Kerr_like_coord} is the standard Kerr spacetime up to $O(b^2)$, see \eqref{metric_prime_Kerr_like_coord_limit} and \eqref{alpha_beta_b_small}, this is guaranteed. %(as is the r.h.s. of the Einstein equations, since scalar field \eqref{scalar_rotated} is $\phi_{\text{ph}} \bigl(u,x^2\bigr) \rightarrow  0$ at $u \rightarrow + \infty$). 
The null energy condition for a null vector $k^\mu$,~e.g., $k^\mu=\bigl(1/A(u,x^2),-1/2, 0, 0\bigr)$, suited to \eqref{metric_prime_Kerr_like_coord},
\begin{align}\label{NEC_rotated}
\small T^\mu_\nu k_\mu k^\nu &= - \frac14 {\Bigl( \phi_{\text{ph}} \bigl(u,x^2\bigr) \Bigr)'_u}^2 \\ \nonumber 
= -&\frac{b^2\bigl(u^2+b^2-a^2\cos^2{x^2}\bigr)^2}{2\Bigl(u^2+\bigl(b-a\cos{x^2}\bigr)^2\Bigr)^2\Bigl(u^2+\bigl(b+a\cos{x^2}\bigr)^2\Bigr)^2} \end{align}
is distinctly violated near an arbitrarily small region and slightly violated, $\sim O(b^2)$, at $u \rightarrow + \infty$, as expected. 

Thereby, the discrepancy in the Einstein equations and a violation of the energy dominance conditions are forced into this fairly small domain, for which parameter $b$ is responsible. Among the known literature examples, only the so-called "eye of the storm" regular rotating black hole \cite{Simpson:2021dyo}, being strictly a model, is similar in a sense of satisfying (or slightly violating, as in our case) the classical energy dominance conditions at infinity for external observers.

\section{Discussion}\label{discussion} 
In spite of the regular black hole models' extreme popularity, only a few exact solutions are still known. To~model realistic physical objects, i.e., to impose axial symmetry, one will face the fact that there has not been elaborated a universal technique to generate a rotating solution from a static one. Most of these approaches are applicable either to vacuum cases, to linearized Einstein's equations (which is the same as representing a metric in the Kerr{--}Schild form), or to exact static solutions with a known physical source (which almost all models of static regular black holes cannot boast of).

We have applied the mainstream approach, the Newman{--}Janis algorithm, to a regular static spacetime sustained by a minimally coupled phantom scalar field. This completely solvable example affords the opportunity to show that, at distances of the regularization parameter's order, we can predict or even conclude nothing due to the Einstein equations' discrepancies. Although coordinate complexification in the Newman{--}Janis spirit leads to a regular Kerr-like black hole, to an external observer, this will be nothing more than standard Kerr's spacetime.

Commonly, the models of regular black holes ad hoc are motivated mainly by phenomenology, with possible observational verification. However, many of them, or even pertubatively slowly rotating solutions in alternative theories, are almost indistinguishable\footnote{Let us point out a recent stationary analytical solution to the semiclassical Einstein equations sourced by the trace anomaly, which can be potentially distinguished from Kerr's solution~\cite{Fernandes:2023vux}.} from the GR solutions from an observational point of view \cite{Psaltis:2007cw,Pani:2009wy,Shaikh:2021yux,Lima:2021las}. Moreover, the cost of simplicity and phenomenological “appealing”, e.g., of the Simpson{--}Visser spacetime, are non-trivial physical sources. 
%Does it really worth to try phenomenological approaches to regular rotating black hole models? 
Is it worth enforcing an exotic matter description for the static regular spacetimes, hereafter imposing axial symmetry approaches to regular rotating black hole models?
Or is the search~for alternative GR singularity treatments more perspective?

%\vspace{-0.5cm} \begin{acknowledgements}  \end{acknowledgements}\vspace{-0.5cm}

\end{document}